\documentclass[%
 reprint,
 amsmath,amssymb,
 aps,
prb,
]{revtex4-2}

\usepackage{graphicx}
\usepackage{dcolumn}
\usepackage{xcolor}%
\usepackage{soul}

\usepackage{braket}

\usepackage{pdfpages} 
\usepackage{pgffor} 

\makeatletter
\AtBeginDocument{\let\LS@rot\@undefined}
\makeatother

\def\supplementfilename{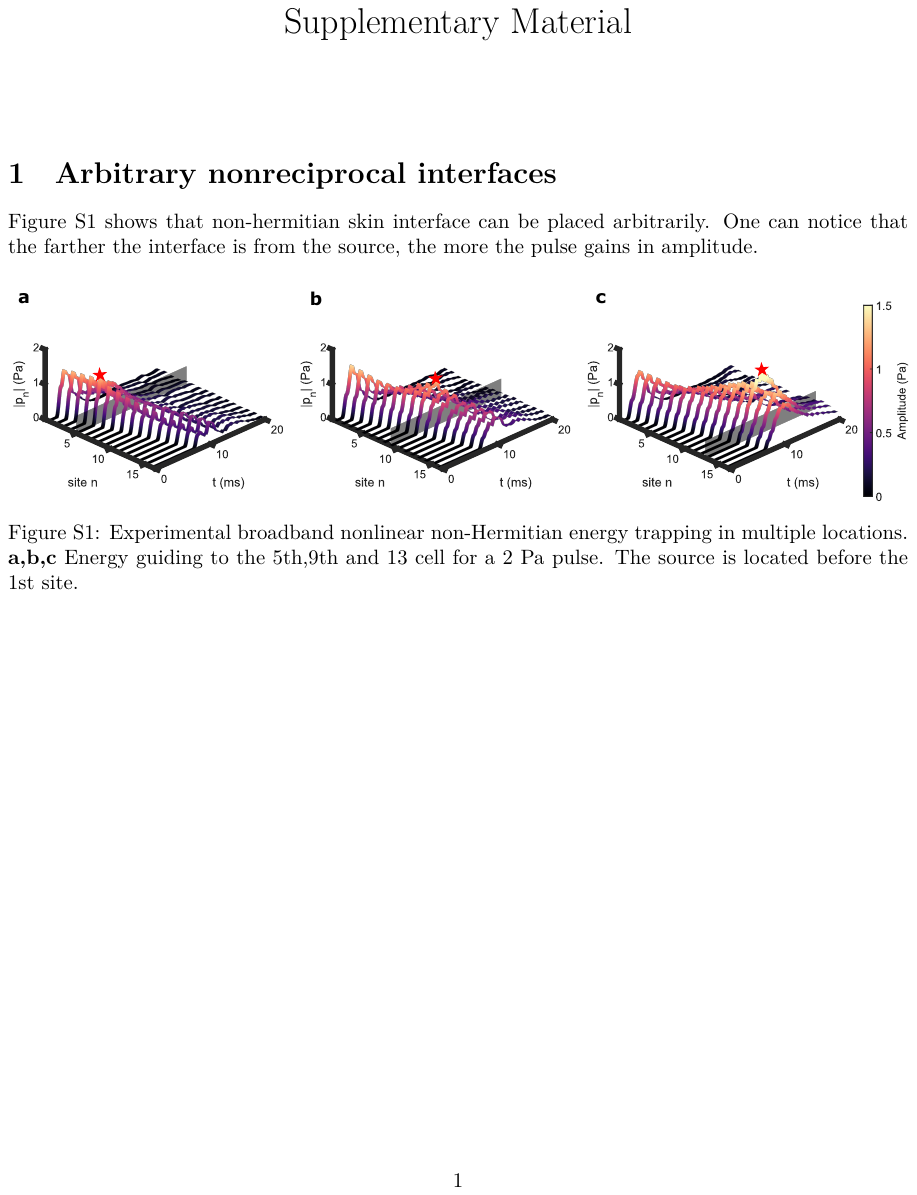}

\pdfximage{\supplementfilename}
\def\numbersupplementpages{\the\pdflastximagepages}

\newif\ifarXiv
\arXivtrue 
\newcommand{\corr}[1]{\textcolor{black}{#1}}

\begin{document}
\preprint{APS/123-QED}

\title{\corr{Observation of} amplitude-driven nonreciprocity for energy guiding}

\author{Mathieu Padlewski}
 \email{mathieu.padlewski@epfl.ch}
 \affiliation{Signal Processing Laboratory 2, Ecole Polytechnique Fédérale de Lausanne, 1015 Lausanne, Switzerland}

\author{Romain Fleury}%
 \email{romain.fleuryk@epfl.ch}
\affiliation{Laboratory of Wave Engineering, Ecole Polytechnique Fédérale de Lausanne, 1015 Lausanne, Switzerland}%

\author{Hervé Lissek}%
 \email{herve.lissek@epfl.ch}
\affiliation{Signal Processing Laboratory 2, Ecole Polytechnique Fédérale de Lausanne, 1015 Lausanne, Switzerland}%

\date{\today}

\begin{abstract}
\corr{The non-Hermitian skin effect is an intriguing physical phenomenon, in which all eigen-modes of a non-Hermitian lattice become localized at boundary regions. While such an exotic behavior has been demonstrated in various physical platforms, most realizations have been so far restricted to the linear regime. Here, we explore the cooperation between nonlinearity and the non-Hermitian skin effect, revealing extraordinary amplitude-driven skin localization dynamics. By introducing an extension to the Hatano-Nelson model where couplings inherent nonlinear behavior, we demonstrate the existence of unique amplitude-driven non-Hermitian skin modes capable of concentrating the energy of a source at any point of space, depending on the power level. Our theoretical model is supported by numerical simulations and experimental realization via a highly configurable acoustic metamaterial composed of active electroacoustic resonators. In all, our findings open up exciting paths for a new generation of non-reciprocal systems, in which nonlinearities serve as a strategic tuning knob to manipulate the guiding process.}
\end{abstract}
\maketitle

\section{Introduction}\label{sec1}

\corr{Hermiticity of a Hamiltonian constitutes a fundamental principle that is integral to various branches of physics. It ensures the conservation of quantities such as energy, momentum, probability amplitude, or electrical charge throughout physical processes. Supported by a well-established formalism, these systems with conserved quantities constitute a significant portion of the research typically conducted by physicists. However, some systems have been shown to exhibit unusual features that fundamentally challenge the conventional constraints imposed by Hermitian systems \cite{bender_real_1998,bender_making_2007,lin_unidirectional_2011,fleury_sound_2014,brandenbourger_non-reciprocal_2019,guo_observation_2023,miri_exceptional_2019,wiersig_prospects_2020,wiersig_review_2020,jiang_interplay_2019,zhu_photonic_2020,weidemann_topological_2020,zhang_observation_2021,brandenbourger_non-reciprocal_2019,ghatak_observation_2020,scheibner_non-hermitian_2020,jiang_interplay_2019,lang_dynamical_2021,zhang_acoustic_2021,gu_transient_2022,maddi_exact_2024} - Non-reciprocal transport \cite{lin_unidirectional_2011,fleury_sound_2014,brandenbourger_non-reciprocal_2019,guo_observation_2023}, exceptional-point-based sensing \cite{miri_exceptional_2019,wiersig_prospects_2020,wiersig_review_2020} and most notably the non-Hermitian skin effect (NHSE) \cite{jiang_interplay_2019,zhu_photonic_2020,weidemann_topological_2020,zhang_observation_2021,brandenbourger_non-reciprocal_2019,ghatak_observation_2020,scheibner_non-hermitian_2020,jiang_interplay_2019,lang_dynamical_2021,zhang_acoustic_2021,gu_transient_2022,maddi_exact_2024} all fall within the broader class of non-Hermitian systems.}

\corr{The NHSE is characterized by the exponential localization of bulk modes near specific boundaries, driven by asymmetric inter-atomic couplings. This phenomenon has garnered significant attention due to recent advancements in applied topology, which have exposed profound topological foundations. In all, these discoveries provide a promising perspective on non-Hermitian phenomena~\cite{okuma_non-hermitian_2023, zhang_review_2022,lin_topological_2023}. This surge of interest in non-Hermitian systems was also swollen by the maturing realm of metamaterials, engineered structures designed to display properties that mimic or even surpass those found in nature.\cite{zheludev_metamaterials_2012, kadic_3d_2019,xiao_active_2020}. In particular, the use of active components has enabled various experimental realizations of NHSE in photonics \cite{zhu_photonic_2020,weidemann_topological_2020}, mechanics \cite{brandenbourger_non-reciprocal_2019,ghatak_observation_2020,scheibner_non-hermitian_2020}, electronics \cite{jiang_interplay_2019,lang_dynamical_2021} and acoustics \cite{zhang_acoustic_2021,maddi_exact_2024}.} 

\corr{Despite the plethora of inquires regarding the interplay between topology and the NHSE , these have been largely addressed in the scope of linear physics. The tendency of describing physical phenomena in linear terms is usually not a manifestation of oversight but rather of convenience - both theoretically and experimentally. However, nonlinearity, when properly tamed, can lead to spectacular applications such as ultra-directional sound sources using parametric arrays~\cite{westervelt_parametric_1963}, energy harvesting of vortices generated by wind turbine blades~\cite{le_fouest_optimal_2024} and deep neural networks~\cite{goodfellow_deep_2016} to name a few. 
Nevertheless, the nonlinear NHSE has seen some theoretical progress for onsite nonlinearities \cite{yuce_nonlinear_2021,ezawa_dynamical_2022,manda_skin_2023}. These local nonlinearities, broadly used in many other topological systems~\cite{zangeneh-nejad_nonlinear_2019,ozawa_topological_2019,smirnova_nonlinear_2020,dobrykh_nonlinear_2018,maczewsky_nonlinearity-induced_2020,hu_nonlinear_2021,chaunsali_self-induced_2019,chaunsali_stability_2021,hadad_self-induced_2016,bisianov_stability_2019,smirnova_topological_2019,tuloup_nonlinearity_2020,kirsch_nonlinear_2021,sohn_topological_2022,manda_wave-packet_2023,chaunsali_dirac_2023}, are typically realized by using saturable components. While ease of realization likely contributes to the large amount of these investigations, saturation unfortunately comes at the cost of being unpredictable and/or complex nonlinear functions. On the other hand, investigations regarding specific nonlocal coupling nonlinearities within the framework of the NHSE remains largely unexplored.}

 \corr{To resolve this shortfall, we extend the NHSE into the nonlinear regime by introducing coupling nonlinearities in a non-Hermitian lattice, revealing a new class of amplitude-driven non-Hermitian skin modes .} \corr{To demonstrate how these modes selectively localize above a critical intensity threshold, we first present an extended version of the Hatano-Nelson model that serves as the foundation for numerical simulations in one- and two- dimensional configurations. Finally, we validate the predictions experimentally via a fully configurable active acoustic metamaterial.} Leveraging these findings, we demonstrate their relevance for selectively focusing energy from high-power sources \corr{toward arbitrary locations}.
\section{Theoretical predictions}\label{sec2}
\subsection{General tight-binding model}\label{sec2}
\begin{figure}
    \centering
    \includegraphics[width=0.4\textwidth]{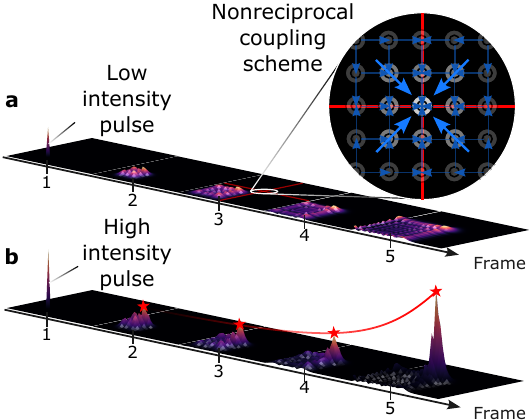}
    \caption{Dynamics of the nonlinear non-Hermitian skin effect in two dimensions. \textbf{a, b} Simulation time frames of the two-dimensional version of the nonlinear model discussed in Fig. 1. The unit cell consists of evanescently-coupled resonators with specified coupling coefficients that are arranged such as to guide the energy to the center (red cross-hair). Furthermore, the non-reciprocal extra-cell coupling coefficients $t^+$, and $t^-$ are nonlinear, enabling an intensity-dependent non-Hermiticity. \textbf{a} Low energy pulse case - typical non-dispersive wave-like behavior. \textbf{b} High energy pulse case - energy converges towards the center.}
    \label{fig:theory_2D}
\end{figure}

Consider two tight-binding Hamiltonians, namely $\hat{H}^+$ and $\hat{H}^-$, defined as:
\corr{
\begin{equation}\label{eq:hamiltonian}
    \begin{gathered}
        \hat{H}^\pm = \sum_i U_i \ket{\psi_i}\bra{\psi_i} \\ + \sum_{i,j} \left(t_{i,j}^{\pm}(I) \ket{\psi_i}\bra{\psi_j}+ t_{i,j}^{\mp}(I) \ket{\psi_j}\bra{\psi_i}\right)
    \end{gathered}
\end{equation}
}
in which $U_i$ corresponds to the onsite interaction at site $i$ and \corr{ $t_{i,j}^{\pm}(I)$ to coupling interactions between sites $i$ and $j$. The system can be viewed as a nonlinear extension of the Hatano-Nelson model~\cite{hatano_localization_1996,hatano_non-hermitian_1998} where non-local coupling interactions are specified to manifest some nonlinear character.  More specifically, they assume the form $t_{i,j}^{\pm}(I) = t_0 \pm \alpha_{i,j}I$ , where $t$ is site independent, $\alpha_{i,j}$ is the non-reciprocal parameter, and $I:=\braket{\psi_j|\psi_j}$ is the field intensity. The tight-binding lattices corresponding to one-dimensional $H^+$ and $H^-$ are represented at the top and bottom of Fig.~\ref{fig:theory_1D} for both low and high intensity configurations respectively and illustrated in two dimensions in the inset of Fig.~\ref{fig:theory_2D}a. The nonreciprocal character of the coupling interactions \corr{$t_{i,j}^+(I) \neq t_{i,j}^-(I)$} combined with nonlinearity triggers an amplitude-driven NHSE, as we demonstrate in the following.}

\subsection{One-dimensional chain}\label{sec2}

\begin{figure}
    \centering
    \includegraphics[width=0.4\textwidth]{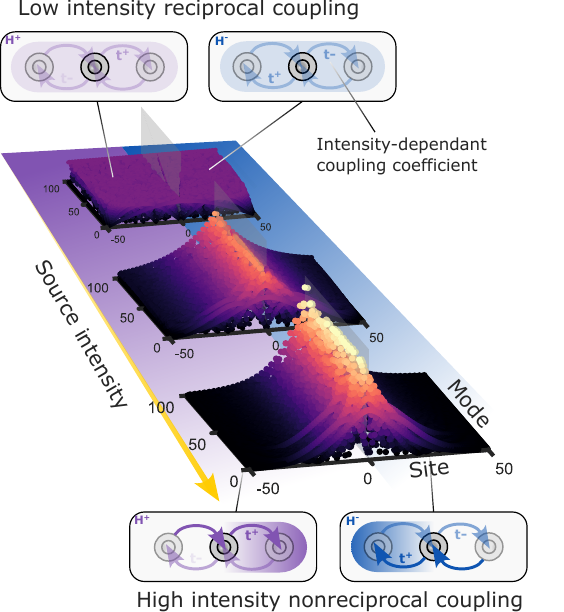}
    \caption{\corr{Nonlinear non-Hermitian skin effect: Eigenmode distribution in a finite chain. Consider a tight-binding lattice composed of resonators coupled to each other with specified hopping strengths. The cell hoppings ($t_+(I)$ and $t_-(I)$) are non-reciprocal and intensity-dependent, which are key ingredients to nucleate a nonlinear non-Hermitian skin effect. The system is partitioned such as to yield an interface between two finite chains (50 unit cells) of inverted coupling lattices. The eigen-mode distributions are plotted for increasing non-Hermiticity ($I\approx0 \rightarrow I=I_{max} $). At low intensity, the eigen-mode spectrum consists of many delocalized states, spreading over the bulk of the lattice. As the intensity level is increased, nonlinear dynamics are triggered causing all eigen-modes to localize at the interface.}}
    \label{fig:theory_1D}
\end{figure}


 Consider a finite chain partitioned in one part described by $H^+$ and another by $H^-$ such that it forms an interface. In other words, the system is configured such that the nonlinear couplings are favored towards the interface. Since the effective Hamiltonians $H^\pm$ are amplitude-dependent, the eigenmodes of the corresponding system are investigated for different excitation energies $I$. Indeed, the nonlinearity of our proposed model is related to the fact that the associated Hamiltonian is a function of its eigenmodes. Therefore, the associated eigenvectors should be calculated throughout a recursive process. In this process, an initial guess is assumed for the eigenvectors. This initial guess is used to generate a new Hamiltonian, based on which a new set of eigenvectors is obtained. This procedure is repeated until achieving a converged solution. In our eigen-mode calculations, we used the relative error of 1\% as the convergence criterion. Our initial guess for the eigen-vector $\ket{\psi}$ was the solution of Eq.~\eqref{eq:hamiltonian} in the absence of nonlinearity (note that for the linear case the eigenvectors of Eq.~\eqref{eq:hamiltonian} can be easily obtained by setting the determinant of the Hamiltonian to zero). The results are presented in Figure~\ref{fig:theory_1D}. For $I\approx 0$, the eigenmode distribution across the chain is uniform, which is reminiscent of a reciprocal Hermitian system. Indeed, the latter is simply a consequence of a vanishingly small nonlinear coupling, i.e. $t_{i,j}^{\pm}(I) \approx t_0$. Increasing the intensity however, unequivocally drives all the modes towards the interface where they localize.

\begin{figure}
    \centering
    \includegraphics[width=0.4\textwidth]{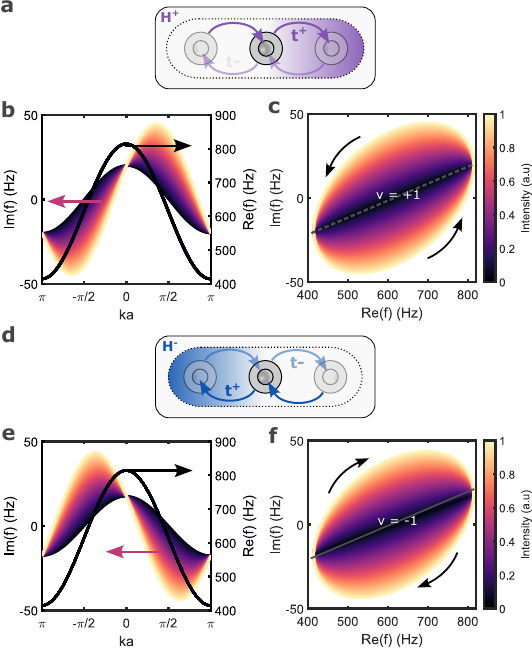}
     \caption{\corr{Topological amplitude-dependant Non-Hermitian skin effect. \textbf{a-c }Right-mover coupling configuration described by $H^+$. \textbf{d-f} Left-mover coupling configuration described by $H^-$. \textbf{a,d} Schematic of the coupling configuration. \textbf{b,e} Real and imaginary parts of the eigenfrequency spectrum calculated with periodic boundary conditions.  \textbf{c,f} The complex eigenfrequency spectrum calculated in \textbf{b,e} forms a closed loop in the complex frequency plane, with its orientation indicating the winding number $\nu$. The dashed line corresponds to the eigenfrequency spectrum calculated with the open boundary conditions.}}
    \label{fig:sup_topo}
\end{figure}

Note that the level of intensity $I$ acts as an effective tuning parameter for the coupling imbalance and does not alter the topological invariant. For concreteness, consider an active phononic crystal described by $H^+$ where a state $\ket{\psi_i} := p_i$ is the complex pressure at site $i$. In other words, the nonlinear couplings are favored towards the right. Imposing periodic boundary conditions reduces Eq.\eqref{eq:hamiltonian} to:
\begin{equation}\label{eq:bloch_hamiltonian}
        \hat{H}^+= \omega_0 + t^+(I)e^{ika} + t^-(I)e^{-ika}
\end{equation}
where $k$ is the wavenumber and $a$ is the lattice spacing and $\omega_0$ is the Floquet-Bloch angular frequency (at the edge of the first Brillouin zone) which effectively operates as the diagonal potential term $U_i$. The system is schematically shown in Figure~\ref{fig:sup_topo}a. The local field intensity $I \propto |p|^2$ (generated by the source) that serves as an effective tuning knob for nonreciprocity: $t^{\pm}(I) = t_0 \pm \alpha I $. Note that in practice, $t_0$ may become complex if active coupling elements are involved, as these elements are prone to introducing latency. Take, for example, a lattice spacing $a = 14 $ cm with locally resonating elements of angular frequency $\omega_0 = 2\pi\cdot614.3$ rad/s and a complex coupling $t_0 = 2\pi\cdot(100 + 8i)$ rad/s. Figure~\ref{fig:sup_topo}b presents the imaginary and real parts of the eigenfrequency spectrum as a function of the local field intensity $I$ for this system. An asymmetric distribution of the imaginary part with respect to $ka=0$ appearing as soon as $I\neq 0$ indicates non-reciprocity. The corresponding eigenfrequency spectrum in the complex frequency plane, plotted in Figure~\ref{fig:sup_topo}c, winds around a tilted elliptical loop, along the direction of increasing wavenumber $ka$ from $0$ to $2\pi$. A quantized integer, known as the winding number $\nu$, can be geometrically obtained by counting the number of loops around some reference frequency $\omega_b$ contained within the complex loop. Alternatively, one can also compute the index directly via \cite{gong_topological_2018, bergholtz_exceptional_2021}:
\begin{equation}
v =\frac{1}{2 \pi i} \int_{-\pi}^\pi \frac{\partial \omega(k) / \partial k}{\omega(k)-\omega_b} d k 
\end{equation}
where $\omega(k)$ is the eigenfrequency band. Either way, here, choosing $\omega_b = 2\pi\cdot600$ yields $\nu_+=+1$. Conversely, the topological index for $H^-$, coupling to the left, is obtained similarly (Fig.~\ref{fig:sup_topo}d-f): $\nu_-=-1$.

It is crucial to notice that the field intensity $I$ has no influence on the topological index $\nu$ as it solely affects the coupling ratio $t^+(I)/t^-(I)$ and not the winding direction. Indeed, the system topology is fixed a priori and the intensity $I$ only accentuates non-reciprocity and consequently, the NHSE. In fact, this relationship between the spectral area and the skin effect lies at the heart of a recent study by K.Zhang \textit{et al.} \cite{zhang_universal_2022}. Thus, the local field intensity $I$ here serves as an effective tuning parameter for the NHSE. 

\subsection{Two-dimensional lattice}\label{sec2}

\begin{figure*}
    \centering
    \includegraphics[width=0.8\textwidth]{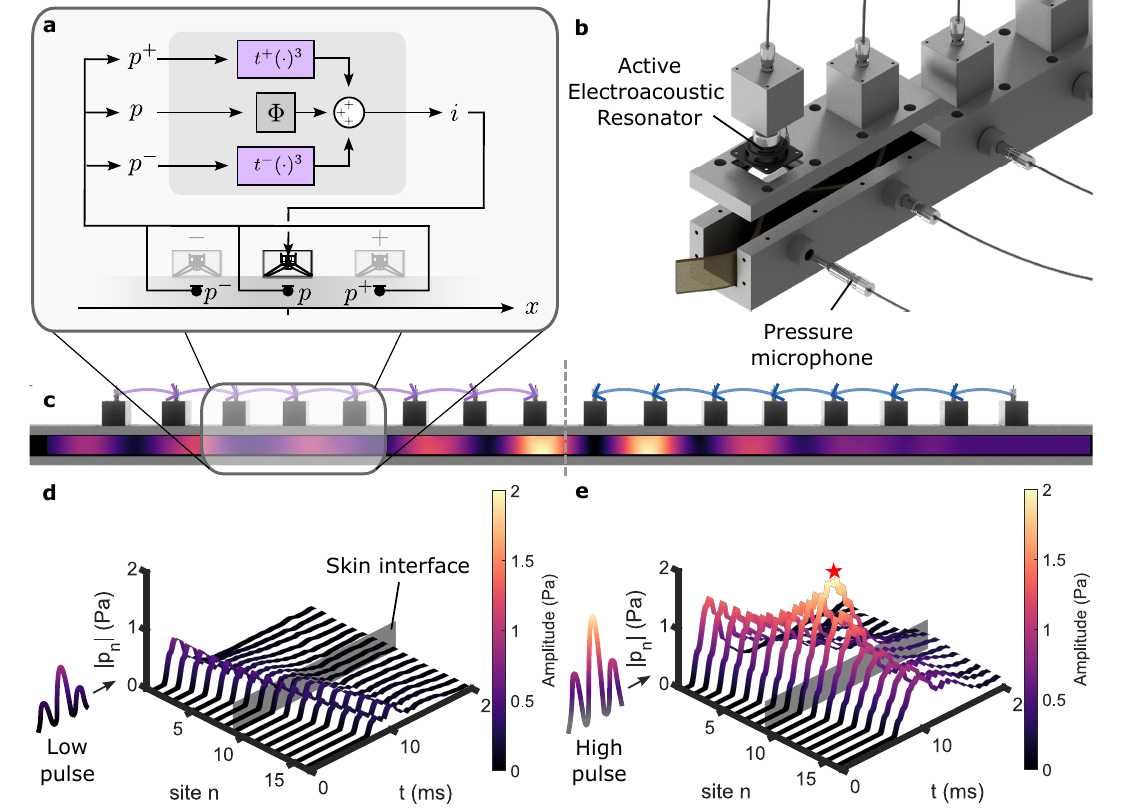}
     \caption{Experimental broadband nonlinear non-Hermitian sound guiding. \textbf{a} Electronic coupling scheme of an active electroacoustic resonator (AER) that depends on the pressure amplitude of the nearest neighbor. The non-reciprocal extra-cell coupling coefficients $t^+$, and $t^-$ are nonlinear, enabling an intensity-dependent non-Hermiticity. \textbf{b} Close-up of an active cell. \textbf{c} Cross-section of the entire metacrystal composed of 16 individually controlled AERs. The waveguide is capped by a source speaker on one end and an anechoic termination on the other. \textbf{d,e} Spatio-temporal plots of a system configured such as to have nonlinear couplings guiding the energy to the center (highlighted in grey) for low and high energy pulse excitations respectively.}
    \label{fig:exp_1D}
\end{figure*}
The remarkable amplitude-dependence of the non-Hermitian skin effect offers a unique opportunity to design exotic systems capable of selectively guiding broadband wave packets to any location in space. Picture a fictional table top composed of atoms that are nonlinearly and nonreciprocally coupled towards its center. A light tap on the surface would evoke usual non-dispersive acoustic waves across the plane whereas, on the other hand, a hard hit would spark nonlinear nonreciprocity and guide the energy toward its center - regardless of the location of the impact. This scenario was executed by solving the time-dependent Schrödinger equation $i\partial_t \ket{\psi}= \hat{H}^{\pm} \ket{\psi}$ using a Runge-Kutta iterative method. A weighted Kronecker-Delta was used as the initial pulse condition near the edge of a square lattice. Depending on the initial intensity, the pulse either diffuses throughout the lattice, or converges with increasing amplitude towards its center. Figure~\label{fig:exp_1D} illustrates the coupling configuration and shows sequential temporal frames that depict the evolution of the system for both initial conditions (- animations can be found in the Supplementary Material).

In the following, we turn this fictional system into reality by presenting a highly configurable active experimental platform capable of producing these nonlinear and non-Hermitian skin effects. \\

\section{Experimental validation}\label{sec2}

Amplitude-dependant interface skin modes are experimentally realized, closely matching theoretical predictions. A finite metacrystal composed of 16 individually controlled active \corr{electroacoustic} resonators (AERs) was successfully developed and manufactured.
A comprehensive overview of the experimental apparatus detailing the electronic coupling scheme, the unit cell and the coupling configuration of the metacrystal is presented in Fig.~\ref{fig:exp_1D}a-c.

The acoustic waveguide is composed of CNC-milled 20mm thick PVC slabs lined with 16 off-the-shelf Visaton FRWS 5 - 8 Ohm drivers, 3D-printed back-cavities, and quarter-inch PCB microphones. The pressure microphones placed at the AERs (Fig.~\ref{fig:exp_1D}b) serve both for interaction control and mode probing. Active control is carried out by a commercial Speedgoat machine equipped with the IO135 module. This Direct Memory Access (DMA) module enables both on-site tunability and low-latency driver control over all 16 channels. COMSOL Multiphysics finite element simulations were performed enabling design optimization and experimental validation for linear coupling configurations. In order to account for manufacturing inconsistencies, the mechanical mass, resistance and compliance of each speaker was characterized following methods described by E. Rivet \cite{Rivet2017a} and actively tuned to match one another. Finally, the quality factor of the resonators where all enhanced by synthetically reducing AER acoustic resistance by 80$\%$.  (details in Supplementary Material). As a last note, both the tuning and reduced resistance are captured by the $\Phi$ gate in Figure~\ref{fig:exp_1D}. 

 The first half of the metacrystal is electronically configured such as to favor nonlinear coupling to the right neighbors and the other to the left. In other words, the metacrystal \corr{is configured to host} amplitude-dependant skin modes at its center (at site  $n = 9$). Note that the interface can be arbitrarily shifted to any site along the chain (See Fig.S1 in the Supplementary Material). Amplitude-dependent third order nonlinear couplings are achieved by means of an active feedback loop.  \corr{Figure~\ref{fig:exp_1D}a illustrates how asymmetric coupling of an AER to its neighbors is achieved. Invoking a real-time controller, the non-local pressures $p\pm$ are independently processed and nonlinearly transformed such as to create the desired coupling characteristics. Note that the local pressure $p$ also undergoes some local processing ($\Phi$ gate in Fig~\ref{fig:exp_1D}a) for the purpose of onsite loss reduction and AER tuning (details regarding this impedance synthesis can be found in the Supplementary Material). Nevertheless, this local processing has no direct consequence on the phenomenon studied here. Finally, an output control current $i$ is channelled to the center AER.} Amplitude-dependency is experimentally demonstrated by the introduction of broadband acoustic pulses at both low and high intensity levels at one end of the chain. As shown in Fig.~\ref{fig:exp_1D}d, a low intensity signal propagates unimpeded throughout the system until it reaches the anechoic termination. Conversely, for a high intensity pulse, the behavior is strikingly different. With the activation of the amplitude-driven nonreciprocal coupling, the signal increases in amplitude up to the skin interface and sharply drops after passing the crest. \corr{Note that while the effect of nonlinearity is simply achieved by increasing the source amplitude, it is not a manifestation of gain saturation. On the contrary, the nonlinearity here is well defined (third order) and the localization of bulk modes at the interface only depends on the arbitrary balance between the electronic coupling strength and the source amplitude - a feat made possible thanks to the broad tunability of the experimental apparatus. }

\section{Conclusion}\label{sec2}
\corr{To summarize,  a synthetic non-Hermitian lattice composed of  asymmetrical and nonlinear couplings was studied, leveraging it to induce an amplitude-driven non-Hermitian skin effect. Theoretically predicted by a nonlinearly-extended Hatano-Nelson model and experimentally achieved by means of a fully programmable active acoustic metacrystal, the resulting amplitude-driven skin modes allow for the realization of a spontaneous energy guiding system, which collects energy from the source during high-power operation and conserves it otherwise.  Although our study focuses on a relatively straightforward nonlinear non-Hermitian system, we anticipate that substantial progress in comprehending the interplay between nonlinearity, non-Hermiticity, and topology can be achieved through the use of highly configurable general purpose metamaterials. Similarly to analogue quantum simulators, the latter could enable rapid and practical realizations of arbitrary synthetic materials, thereby facilitating deeper inquiries into its complex interactions.}\\

\textbf{Acknowledgements} This research was supported by the Swiss National Science Foundation under Grant No. $200020 \_ 200498$.
\bibliographystyle{apsrev4-1}
\bibliography{main_bibliography.bib}

\ifarXiv
    \foreach \x in {1,...,\numbersupplementpages}
    {
        \clearpage
        \includepdf[pages={\x,{}}]{\supplementfilename}
    }
\fi

\end{document}


\maketitle
\beginsupplement

 \section{Arbitrary nonreciprocal interfaces}
\begin{figure}[H]
    \centering
    \includegraphics[width=1\textwidth]{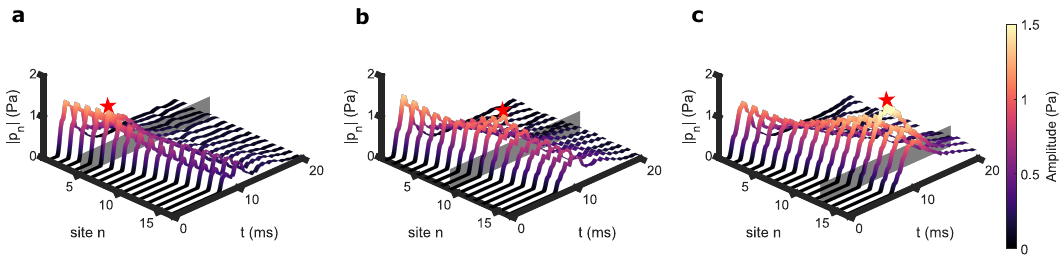}
     \caption{Experimental broadband nonlinear non-Hermitian energy trapping in multiple locations. \textbf{a,b,c} Energy guiding to the 5th, 9th and 13 cell for a 2 Pa pulse. The source is located before the 1st site.}
    \label{fig:sup_exp_1D}
\end{figure}

Figure~\ref{fig:sup_exp_1D} shows that non-Hermitian skin interface can be placed arbitrarily. One can notice that the farther the interface is from the source, the more the pulse gains in amplitude.

\section{Temporal dynamics simulation of a nonlinear and nonreciprocal system using a lumped parameter model }
\begin{figure}[H]
    \centering
    \includegraphics[width=1\textwidth]{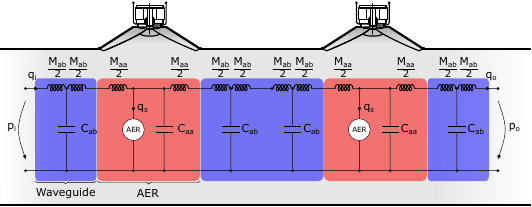}
    \caption{Scattering dynamics through a unit cell using acoustic circuit formalism. The latter specifies the relation between input and output pressures and volumetric flows, $(p_i,q_i)$ and $(p_o,q_o)$ respectively. Three waveguide parts and two AER parts, highlighted in blue and red respectively, are placed within the corresponding experimental unit cell schematically drawn in grey. }
    \label{fig:theory_dynamic_matrix}
\end{figure}

The aim here is to build a dynamical matrix corresponding to a $N$-cell acoustic scattering chain of electronically controlled resonators in view of faithfully predicting the temporal evolution of nonlinear topological modes. Local resonating dynamics can elegantly be captured by means of an analogous circuit formalism where electrical potential and current are analogically substituted with acoustic pressure $p$ and volumetric flow $q$. The electrical inductance, resistance and capacitance are respectively changed to the acoustic mass $M_a$, resistance $R_a$ and compliance $C_a$. Figure~\ref{fig:theory_dynamic_matrix} depicts the circuit describing an acoustic unit cell composed of a waveguide loaded with two equally spaced AERs.  The circuit is partitioned into three identical waveguide and two AER parts - each of which is defined by a sub-dynamical matrix which will be elaborated below. The product of all latter five elements yields the total unit cell dynamical matrix. Table~\ref{tab:parameters} summarizes the values used for both theoretical and experimental approaches.    \\ 

\begin{table}[h]
	\centering
    \caption{Experimental and Simulation Parameters}\label{tab:parameters}
    \begin{tabular}{lccc}
    \hline
    Parameter                     & Symbol  & Value & Unit \\ \hline
    Ac. Mass between AERs        & $M_{ab}$  & $29.0$ & $\si{ kg/m^4}$ \\ 
    Ac. Compliance between AERs  & $C_{ab}$  & $7.2\cdot10^{-10}$& $\si{m^3/Pa}$\\ 
    Ac. Mass at AER             & $M_{aa}$  & $23.9$& $\si{ kg/m^4}$\\ 
    Ac. Compliance at AER       & $C_{aa}$  & $5.9\cdot10^{-10}$& $\si{m^3/Pa}$ \\ 
    AER Ac. Mass                & $M_{as}$  & $4.5\cdot10^{4}$& $\si{ kg/m^4}$ \\ 
    AER Ac. Resistance          & $R_{as}$  & $463.7$& $\si{Pa s/m}$ \\ 
    AER Ac. Compliance          & $C_{as}$  & $3.1\cdot10^{-10}$ & $\si{m^3/Pa}$ \\ 
    Force Factor                & $B l$     & $1.4$& $\si{Pa/A}$ \\ 
    Effective Diaphragm Area    & $S_d$     & $12\cdot10^{-4}$& $\si{m^2}$ \\ 
    \end{tabular}
\end{table}

\textbf{Waveguide part:}
The relation between the input and output pressures and volumetric flows in a simple waveguide section, $(p_i,q_i)$ and $(p_o,q_o)$ respectively, is given by the following dynamic equations:
\begin{gather*}
    \Longrightarrow
    \left \{
    \begin{split}
    -p_i = -\frac{M_{ab}}{2}\frac{dq_i}{dt} - \frac{1}{C_{ab}}\int  (q_i-q_o) \,dt \\
    \frac{1}{C_{ab}}\int  (q_i-q_o) - \frac{M_{ab}}{2}\frac{dq_o}{dt} = p_o
    \end{split}
    \right .
\end{gather*}

The latter can be conveniently written in matrix form:
\begin{gather*}
    \Rightarrow
    \left \{
    \left[
    \color{\colorb}
    \begin{matrix}
        -\frac{M_{ab}}{2} & 0 \\
        0 & -\frac{M_{ab}}{2}
    \end{matrix}
    \color{black}
    \right] 
    \frac{d^2}{dt^2}
    +
    \left[
    \color{\colorb}
    \begin{matrix}
        -C_{ab}^{-1} & C_{ab}^{-1} \\
        C_{ab}^{-1} & -C_{ab}^{-1}
    \end{matrix}
     \color{black}
    \right]
    \right \}
    \begin{bmatrix}
        x_i\\
        x_o
    \end{bmatrix}
    =	
    \begin{bmatrix}
        -p_i\\
        p_o
    \end{bmatrix}
\end{gather*}

\textbf{Active Electro-Acoustic Resonator part:}
Similarly to the waveguide, the relation between $(p_i,q_i)$ and $(p_o,q_o)$ is given by the following dynamic equation:
\begin{gather*}
    \Longrightarrow
     \left \{
    \begin{split}
        -p_i = -\frac{M_{aa}}{2}\frac{dq_i}{dt} - \frac{1}{C_{aa}}\int  (q_i-q_s-q_o) \,dt \\
        -\hat{\zeta}q_s -\frac{Bli}{S_d} - \frac{1}{C_{aa}}\int  (q_i-q_s-q_o) \,dt  = 0\\
        \frac{1}{C_{aa}}\int  (q_i-q_s-q_o) \,dt  = \frac{M_{aa}}{2}\frac{dq_o}{dt} + p_o
    \end{split}
    \right .
\end{gather*}

where the resonator dynamics are captured by the AER impedance operator,
	\begin{equation*}
		\hat{\zeta} = M_{as}\frac{d}{dt} + R_{as} + \frac{1}{C_{aa}}\int \,dt
	\end{equation*}
and the controllable electrical current $i$ is at the root of a supplementary resonator interactions - local and nonlocal.

In matrix form, the latter yields:
\begin{gather*}
    \Rightarrow
    \left \{
    \left[
    \color{\colora}
    \begin{matrix}
        -\frac{M_{aa}}{2}& 0 & 0 \\
        0 & M_{as} & 0 \\
        0 & 0 & -\frac{M_{aa}}{2}
    \end{matrix}
    \color{black}
    \right]
    \frac{d^2}{dt^2}
    +
    \left[
    \color{\colora}
    \begin{matrix}
        0 & 0 & 0 \\
        0 & R_{as} & 0 \\
        0 & 0 & 0
    \end{matrix}
    \color{black}
    \right]
    \frac{d}{dt}
    +
    \left[
    \color{\colora}
    \begin{matrix}
        -C_{aa}^{-1} & C_{aa}^{-1} & C_{aa}^{-1} \\
        -C_{aa}^{-1} & C_{as}^{-1}+C_{aa}^{-1} & C_{aa}^{-1} \\
        C_{aa}^{-1} & -C_{aa}^{-1} & -C_{aa}^{-1}
    \end{matrix}
    \color{black}
    \right]
    \right \}
    \begin{bmatrix}
        x_i\\
        x_s\\
        x_o
    \end{bmatrix}
    = \\
    \begin{bmatrix}
        -p_i\\
        0\\
        p_o
    \end{bmatrix}
    +
    \begin{bmatrix}
        0\\
        -\frac{Bli}{S_d}\\
        0
    \end{bmatrix}
\end{gather*}

    \textbf{Unit cell dynamical matrix:}\\
 The evolution of an acoustic charge $\mathbf{x} \coloneqq \int \mathbf{q} \,dt$ at each circuit node in Figure~\ref{fig:theory_dynamic_matrix} is computed using the following dynamical system where the colored block-diagonal matrices correspond to those used in the aforementioned waveguide and AER:
    \begingroup
    \begin{gather*}
		\left \{
		\renewcommand{\arraystretch}{1} 
		\setlength{\arraycolsep}{1pt} 
		\underbrace{
        \tiny
		\begin{bmatrix}
			\color{\colorb}\bullet & \color{\colorb}\bullet &   &  &  &  &  &  &  \\
			\color{\colorb}\bullet & \color{\colorab}\bullet & \color{\colora}\bullet & \color{\colora}\bullet &  &  &  &  &  \\
			& \color{\colora}\bullet & \color{\colora}\bullet & \color{\colora}\bullet &  &  &  &  &  \\
			& \color{\colora}\bullet & \color{\colora}\bullet & \color{\colorab}\bullet & \color{\colorb}\bullet &  &  &  &  \\
			&  &  & \color{\colorb}\bullet & \color{\colorb}\bullet & \color{\colorb}\bullet &   &  &  \\
			&  &  &  & \color{\colorb}\bullet & \color{\colorab}\bullet & \color{\colora}\bullet & \color{\colora}\bullet &  \\
			&  &  &  &   & \color{\colora}\bullet & \color{\colora}\bullet & \color{\colora}\bullet &  \\
			&  &  &  &  & \color{\colora}\bullet & \color{\colora}\bullet & \color{\colorab}\bullet & \color{\colorb}\bullet \\
			&  &  &  &  &  &  & \color{\colorb}\bullet & \color{\colorb}\bullet
		\end{bmatrix}
		}_{\normalfont {M_\text{cell}}}
		\frac{d^2}{dt^2}
		+
		\underbrace{
         \tiny
		\begin{bmatrix}
			\color{\colorb}\bullet & \color{\colorb}\bullet &   &  &  &  &  &  &  \\
			\color{\colorb}\bullet & \color{\colorab}\bullet & \color{\colora}\bullet & \color{\colora}\bullet &  &  &  &  &  \\
			& \color{\colora}\bullet & \color{\colora}\bullet & \color{\colora}\bullet &  &  &  &  &  \\
			& \color{\colora}\bullet & \color{\colora}\bullet & \color{\colorab}\bullet & \color{\colorb}\bullet &  &  &  &  \\
			&  &  & \color{\colorb}\bullet & \color{\colorb}\bullet & \color{\colorb}\bullet &   &  &  \\
			&  &  &  & \color{\colorb}\bullet & \color{\colorab}\bullet & \color{\colora}\bullet & \color{\colora}\bullet &  \\
			&  &  &  &   & \color{\colora}\bullet & \color{\colora}\bullet & \color{\colora}\bullet &  \\
			&  &  &  &  & \color{\colora}\bullet & \color{\colora}\bullet & \color{\colorab}\bullet & \color{\colorb}\bullet \\
			&  &  &  &  &  &  & \color{\colorb}\bullet & \color{\colorb}\bullet
		\end{bmatrix}
		}_{\normalfont {R_\text{cell}}}
		\frac{d}{dt}
		+
		\underbrace{
        \tiny
		\begin{bmatrix}
			\color{\colorb}\bullet & \color{\colorb}\bullet &   &  &  &  &  &  &  \\
			\color{\colorb}\bullet & \color{\colorab}\bullet & \color{\colora}\bullet & \color{\colora}\bullet &  &  &  &  &  \\
			& \color{\colora}\bullet & \color{\colora}\bullet & \color{\colora}\bullet &  &  &  &  &  \\
			& \color{\colora}\bullet & \color{\colora}\bullet & \color{\colorab}\bullet & \color{\colorb}\bullet &  &  &  &  \\
			&  &  & \color{\colorb}\bullet & \color{\colorb}\bullet & \color{\colorb}\bullet &   &  &  \\
			&  &  &  & \color{\colorb}\bullet & \color{\colorab}\bullet & \color{\colora}\bullet & \color{\colora}\bullet &  \\
			&  &  &  &   & \color{\colora}\bullet & \color{\colora}\bullet & \color{\colora}\bullet &  \\
			&  &  &  &  & \color{\colora}\bullet & \color{\colora}\bullet & \color{\colorab}\bullet & \color{\colorb}\bullet \\
			&  &  &  &  &  &  & \color{\colorb}\bullet & \color{\colorb}\bullet
		\end{bmatrix}
		}_{\normalfont {K_\text{cell}}}
		\right \}
		\underbrace{
		\begin{bmatrix}
			x_1 \\
			x_2 \\
			\color{\colora}{{x}_A}\\
			x_4 \\
			x_5 \\
			x_6 \\
			\color{\colora}{{x}_B} \\
			x_8\\
			x_9 
		\end{bmatrix} 
		}_{\normalfont {\mathbf{x}}}
		+
		Bl
		\underbrace{
		\begin{bmatrix}
			0 \\
			0 \\
			\color{\colora}{{i}_A} \\
			0 \\
			0 \\
			0 \\
			\color{\colora}{{i}_B}\\
			0 \\
			0 
		\end{bmatrix}
		}_{\normalfont {\mathbf{i}}}
		=
		\underbrace{
		\begin{bmatrix}
			-p_1 \\
			0 \\
			0 \\
			0 \\
			0 \\
			0 \\
			0 \\
			0 \\
			p_9 
		\end{bmatrix}
		}_{\normalfont {\mathbf{p}}}
	\end{gather*}
    \endgroup

    The AER control current $i$ used for generating nearest neighbor coupling is defined as:
    \begin{equation}\label{eq:control_current}
    \Rightarrow \textcolor{\colora}{i_{A(B)}} = \frac{S_d}{Bl} \sum_{\mu}c_\mu (p_{B(A)})^\mu
    \end{equation}
where $ c_\mu$ are the non-linear coupling coefficients of order $\mu$.   

	
	The pressure at the AER is obtained by assuming continuity of pressure:
    \begin{equation}   
        \begin{aligned}
            p_A &= \frac{1}{C_{aa}}(x_{2} - x_{A} - x_{4})\\
            p_B &= \frac{1}{C_{aa}}(x_{6} - x_{B} - x_{8})
        \end{aligned}
    \end{equation}

\begin{figure}[H]
    \centering
    \includegraphics[width=1\textwidth]{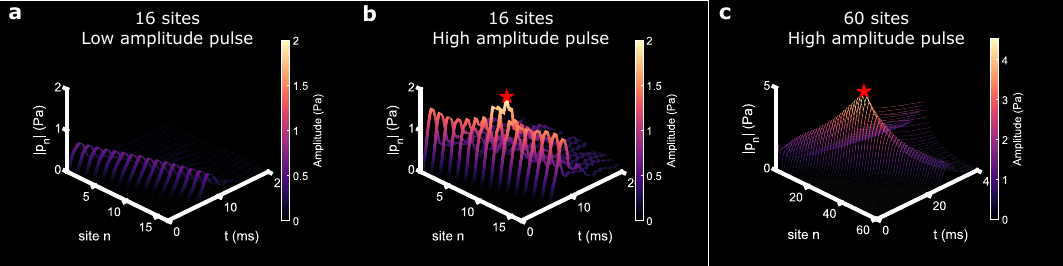}
     \caption{Simulation of broadband nonlinear non-Hermitian energy guiding using the acoustic circuit model. \textbf{a,b} Energy guiding in a 16 site chain for a low and high pulse (1 Pa and 2 Pa resp.) \textbf{c} Energy guiding in a 60 site chain for a high pulse (2 Pa).}
    \label{fig:sup_sim_1D}
\end{figure}

    \textbf{Crystal dynamical matrix:}\\
    Finally, an $N$-cell metacrystal dynamical matrix is built by simply concatenating the unit cell dynamical matrices $M_{\text{cell}},R_{\text{cell}},C_{\text{cell}}$,
    \renewcommand{\arraystretch}{0.65} 
	\setlength{\arraycolsep}{1.5pt} 
    \begin{equation*}\setstretch{1}
        \underbrace{
                \left[
                \quad
                \begin{NiceArray}{ccccccccccccccccccc}
                        &  &   &  &  &  &  &  &  &  &  &  &  &  &  &  &  &  &  \\
                        \color{\colorb}\transparent{0.1} \bullet& \color{\colorb}\transparent{0.25} \bullet   &   &  &  &  &  &  &  &  &  &  &  &  &  &  &  &  &  \\
                        \color{\colorb}\transparent{0.25} \bullet& \color{\colorb}\transparent{0.5} \bullet& \color{\colorb}\bullet &  &  &  &  &  &  &  &  &  &  &  &  &  &  &  &  \\
                        & \color{\colorb}\bullet & \color{\colorab}\bullet & \color{\colora}\bullet & \color{\colora}\bullet &  &  &  &  &  &  &  &  &  &  &  &  &  &  \\
                        &  & \color{\colora}\bullet & \color{\colora}\bullet & \color{\colora}\bullet &  &  &  &  &  &  &  &  &  &  &  &  &  &  \\
                        &  & \color{\colora}\bullet & \color{\colora}\bullet & \color{\colorab}\bullet & \color{\colorb}\bullet &  &  &  &  &  &  &  &  &  &  &  &  &  \\
                        &  &  &  & \color{\colorb}\bullet & \tikzmark{left} \color{\colorb}\bullet & \color{\colorb}\bullet &  &  &  &  &  &  &  &  &  &  &  &  \\
                        &  &  &  &  & \color{\colorb}\bullet & \color{\colorab}\bullet & \color{\colora}\bullet & \color{\colora}\bullet &  &  &  &  &  &  &  &  &  &  \\
                        &  &  &  &  &  & \color{\colora}\bullet & \color{\colora}\bullet & \color{\colora}\bullet &  &  &  &  &  &  &  &  &  &  \\
                        &  &  &  &  &  & \color{\colora}\bullet & \color{\colora}\bullet & \color{\colorab}\bullet & \color{\colorb}\bullet &  &  &  &  &  &  &  &  &  \\
                        &  &  &  &  &  &  &  & \color{\colorb}\bullet & \color{\colorb}\bullet & \color{\colorb}\bullet &  &  &  &  &  &  &  &  \\
                        &  &  &  &  &  &  &  &  & \color{\colorb}\bullet & \color{\colorab}\bullet & \color{\colora}\bullet & \color{\colora}\bullet &  &  &  &  &  &  \\
                        &  &  &  &  &  &  &  &  &  & \color{\colora}\bullet & \color{\colora}\bullet  & \color{\colora}\bullet &  &  &  &  &  &  \\
                        &  &  &  &  &  &  &  &  &  & \color{\colora}\bullet & \color{\colora}\bullet & \color{\colorab}\bullet & \color{\colorb}\bullet &   &  &  &  &  \\
                        &  &  &  &  &  &  &  &  &  &  &  & \color{\colorb}\bullet & \color{\colorb}\bullet \tikzmark{right} & \color{\colorb}\bullet &   &  &  &  \\
                        &  &  &  &  &  &  &  &  &  &  &  &  & \color{\colorb}\bullet & \color{\colorab}\bullet & \color{\colora}\bullet & \color{\colora}\bullet &  &  \\
                        &  &  &  &  &  &  &  &  &  &  &  &  &  & \color{\colora}\bullet & \color{\colora}\bullet  & \color{\colora}\bullet &  &  \\
                        &  &  &  &  &  &  &  &  &  &  &  &  &  & \color{\colora}\bullet & \color{\colora}\bullet & \color{\colorb}\bullet & \color{\colorb}\bullet &  \\
                        &  &  &  &  &  &  &  &  &  &  &  &  &  &  &  & \color{\colorb}\bullet & \color{\colorb}\transparent{0.5} \bullet & \color{\colorb}\transparent{0.25} \bullet   \\
                        &  &  &  &  &  &  &  &  &  &  &  &  &  &  &  &  & \color{\colorb}\transparent{0.25} \bullet   & \color{\colorb}\transparent{0.1} \bullet  \\
                        &  &   &  &  &  &  &  &  &  &  &  &  &  &  &  &  &  &  
                    \end{NiceArray}
                \quad
                \right ]
            }_{(8N+1) \times (8N+1)}	
       \DrawBox[thick, black,  dotted]{left}{right}{\textcolor{black}{\scriptsize unit cell}}
    \end{equation*}
and numerical simulations can be straightforwardly carried out by Runge-Kutta iterative methods such as those presented in Figure~\ref{fig:sup_sim_1D} which closely match the experimental findings presented in the main manuscript.

\section{Impedance synthesis}\label{app:impedance_synth} 

The reason for altering the local response of each AER is twofold: the first is to tune all the resonators to the same frequency such as to have identical "meta-atoms" and the second to reduce onsite losses. The synthesis procedure presented here summarizes the work reported by E. Rivet \textit{et al.} \cite{Rivet2017}.
\begin{figure}[H]
    \centering
	\includegraphics[width=0.25\linewidth]{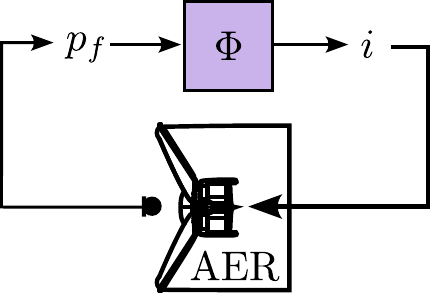}
	\caption{Impedance synthesis control scheme. The front pressure is measured at the front of the AER diaphragm (cross-section) with a microphone and is multiplied by a transfer function $\Phi$ which in turn outputs the drive current required to achieve a specified target impedance. }
	\label{fig:impedance_synth}%
\end{figure}
Let's consider a closed-box electrodynamic loudspeaker subject to an exogenous sound pressure $p_f$ at the diaphragm, and an electrical current $i$ at its electrical terminals as shown in figure~\ref{fig:impedance_synth}. Assuming steady-state and expressed in the frequency domain, its velocity response $v(s)$ is derived from Newton's second law of motion:

	\begin{equation}
		\zeta_{mc}(s)v(s) = S_d p_f(s) - B\ell i(s)
	\end{equation}
	
	Where:
	\begin{itemize}
		\item $\zeta_{mc}$: total mechanical impedance (N.s/m), including the compliance of air inside the cabinet
		\item $v$: diaphragm velocity (m/s)
		\item $S_d$: diaphragm surface area ($m^2$)
		\item $p_f$: pressure in front of diaphragm (Pa)
		\item $B\ell$: Force factor (T.m)
		\item $i$: electrical current (A)
		\item $s = j \omega$: the Laplace variable (Rad/s)
	\end{itemize}

	This allows deriving the electric current $i(s)$ required to achieve a target mechanical impedance $\zeta_{st}$ (Pa.s/m) at the diaphragm:
	
	\begin{equation}
		i(s) = \frac{(\zeta_{st}(s)-\zeta_{mc}(s))\cdot v(s)}{Bl}
	\end{equation}
	
	We can specify 	$\zeta_{st}$ to have the following form:
	
	\begin{equation}
		\zeta_{st}(s) = \mu_M M_{ms}\cdot s+\mu_R R_{ms} + \mu_C/(C_{mc}\cdot s)
	\end{equation}
	
	Where:
	\begin{itemize}
		\item $M_{ms}$: diaphragm mass (kg)
		\item $R_{ms}$: mechanical resistance  
		\item $C_{mc}$: speaker + cabinet compliance ($s^2$/kg)
		\item $\mu_M,\mu_R,\mu_C$: control parameters
	\end{itemize}
	Thus, the control current required to synthesize a target virtual impedance $\zeta_{st}$ is:
	
	\begin{equation}
		i(s) = p_f(s)\cdot\Phi(s)
	\end{equation}
	Where
		\begin{equation} 
			\Phi = \frac{S_d}{Bl}\left[\frac{(\mu_M-1)M_{ms}\cdot s+(\mu_R-1)R_{ms} + (\mu_C-1)/(C_{mc}\cdot s)}{\mu_M M_{ms}\cdot s+\mu_R R_{ms} + 	\mu_C/(C_{mc}\cdot s)}\right]
		\end{equation}
	
	is the transfer function.
	
	 In practice, sound waves arriving at the microphones are detected, digitally sampled, and processed in real-time by a software program running on a commercial Speedgoat machine equipped with the IO135 module. The digital output signals are then converted back to analog and are fed back to the speakers to generate the desired synthesized response. Direct Memory Access allows for processing delays as little as 25 $\mu s$ which enables large synthesis as explained in a recent stability study carried out by De Bono \textit{et al.}\cite{DeBono2022}.

\section{Bibliography}
\bibliographystyle{plain}
\bibliography{./supplementary/supp}